# Electronic and Optical Properties of Zinc based Hybrid Organic-Inorganic Compounds for Ultraviolet Applications

Trilochan Sahoo,[1,*] Uchenna A. Anene,[2] S. Pamir Alpay,[2,3] and Sanjeev K. Nayak [2]

[1] *Department of Physics and Nanotechnology, SRM Institute of Science and Technology, Kattankulathur, 603203, Tamilnadu, India*

[2] *Department of Materials Science and Engineering and Institute of Materials Science University of Connecticut, Storrs, CT 06269, USA*

[3] *Department of Physics, University of Connecticut, Storrs, CT 06269, USA*

## ABSTRACT

Hybrid organic-inorganic (HOI) compounds are excellent candidates for a wide spectrum of applications in diverse fields such as optics, electronics, energy and biotechnology. Their broad range of versatility is achieved by combining the functionalities of organic and inorganic materials to generate unique properties. Current research has mostly focused on perovskite HOIs due to their wide range of uses in solar cells, photo detectors and memory devices. However, drawbacks such as instability and lead toxicity limit further implementation into other new areas. Thus, there is a need to develop stable and non-toxic HOI perovskite materials. Zinc is an attractive substitute for Pb in HOIs. Here, we apply a functionality based materials selection approach to screen for Zn-based HOI compounds from two crystallographic repositories; Inorganic Crystal Structure Database and American Mineralogist Crystal Structure Database. We successfully identify thirteen Zn-based HOI compounds. The electronic structure and optical properties of these compounds are investigated using density functional theory. The calculated optical absorbance fall within the far ultra-violet (FUV) region of 200–112 nm wavelength. We selected four of these compounds and calculated their band gaps; they were found to range between 4.9–5.7 eV. Considering that the UV absorbance is three times larger than average tissue absorbance and the refractive index (> 1.49) is greater than typical tissue materials, one could consider these Zn-based HOI compounds for selective photothermolysis treatment and UV protectant coating for electronic devices.

---

[*] Corresponding author e-mail: trilochs@srmist.edu.in

**Keywords:** Zinc based hybrid organic-inorganic materials; complex oxides; wide band gap; UV application; first principles calculations; computational materials discovery

## 1. Introduction

Hybrid organic and inorganic materials offer the ability to tailor brand-new functionalities by adjusting composition on the atomic/molecular scale. Such a combination of organic–inorganic character provides an opportunity to design interesting properties from the ground up and thereby the realization of a new group of materials specifically conceived for many innovative applications in extremely diverse fields ranging from optics, electronics, photovoltaics, energy storage, and catalysis[1–5]. The wide chemical space of hybrid materials is often accompanied by rich structural forms and polymorphism. This increases adjustability of the material for desired physicochemical properties. The dynamic properties of HOI materials are related to the nature of the bonding between the organic and the inorganic components. HOIs are subclassified into two categories: class I and class II. Weak van der Waals, ionic or hydrogen bonding link the organic and inorganic interfaces of class I hybrids. Class II HOI materials are associated with strong covalent or iono-covalent chemical bonds [6]. Therefore, class II materials are of interest in the context of this study we have carried out in exploration of functional materials for UV-applications.

Due to their versatility, the majority of HOI research is focused on perovskite materials, with general chemical formula $ABO_3$. The A site is typically occupied by a monovalent organic cation, while the B site usually consists of an element from group 14 of the periodic table. The larger cations such as tin ($Sn^{2+}$) and lead ($Pb^{2+}$) allow for the accommodation of organic cations at the A site leading to a hybrid crystal [6–8]. Common examples include, $CH_3NH_3PbI_3$ and $CH_3NH_3PbBr_3$, which are actively investigated for their photovoltaic properties [9]. HOIs have been stabilized with various organic functional groups, such as halides, formates, dyanides, dicyanamides, amides and dicyanometallates [1,10,11]. Ferroelectric properties have been explored in $[(CH_3)_2NH_2][M(HCOO)_3]$ with M = Mg, Fe, Co, Ni, Cu, or Zn, where electric polarization in the range 0.6 μC/cm² – 6 μC/cm² have been observed [12–16]. The magnitude of polarization is not as high as prototypical ferroelectric oxides, e.g., 50 µC/cm$^2$ and 66 µC/cm$^2$ for Bi$_4$Ti$_3$O$_{12}$ and PbTiO$_3$, respectively[17,18], yet might be promising for bio-imaging applications.

More importantly, environmental and sustainability issues motivate the reduction of elements such as Pb from technologies and products. Many HOI perovskites have Pb as a constituent, which is toxic to the environment and poses a health risk to living organisms. There is a drive to remove poisonous materials from the economy chain through design of safe, yet smart materials substitutions [19,20]. Furthermore, successful integration of organic blocks with a nontoxic cation would increase the applicability of HOI materials for a wider variety of uses in areas such as life sciences and medicine. Thus, the need to develop stable and non-toxic HOI perovskite materials is great.

Zinc is a widely used biocompatible element; its cation Zn$^{2+}$ offers an attractive substitute for Pb$^{2+}$ in HOIs. It is a component of many bio-enzymes and plays an important role in metabolizing nucleic acids, thus assisting in cell replication, tissue repair and growth in living organisms[21,22]. The simplest inorganic oxide of Zn, ZnO, is an intriguing material due to its large band gap, high excitonic binding energy, high optical absorbance for ultraviolet (UV) light, and high chemical stability [23–25]. ZnO finds applications in light emitting diodes, optical sensors, piezoelectric devices for energy harvesting and in the biomedical field, for such uses as bioimaging, drug delivery and biosensors[26]. Recently reported stable ZnO monolayers obtained from a mild hydrothermal technique using commonly available precursors of zinc nitrate hexahydrate, hexamethelene tetramine and sodium citrate have opened a new frontier for large scale production of graphene-like ZnO [27]. The two-dimensional ZnO flakes are attractive for nanoelectronics and UV photo detectors [28,29]. Amalgamating organic units and inorganic oxides into a single system would considerably expand the use of these materials use towards bio-applications while the functionalities of inorganic components could still be preserved. This further emphasizes the feasibility for the development of Zn-based HOI materials.

Until now only a few Zn-based HOI compounds have been developed. The discovery of new materials and their development has typically required a bit of intuition, chance, and extensive experimental work. This process takes a relatively long time, ranging from only a couple of years up to decades depending on the specific industry. Programs such as the Materials Genome Initiative (MGI) aim to accelerate materials discovery and development using a combination of techniques including data analysis/curation, computational methodologies at multiple length scales, and judicious experimental work [30–34]. In line with some of these ideas, we present here an efficient and targeted approach that could result in the discovery of stable Zn-based Pb-free HOI perovskite materials. Specifically, we analyze existing experimental databases and perform functionality-based materials screening. We then study the structural, electronic, and optical properties of qualified materials using first principles approaches. Our results show that there are thirteen practically synthesized Zn based HOIs out of which several could potentially lend themselves to UV protection.

## 2. Computational Methodology

The physical properties of the compounds are obtained from electronic structure calculations based on DFT. The plane wave pseudopotential approach implemented in the Vienna *ab initio* simulation package (VASP) [35,36] was used. The screened hybrid functional (SHF) method as proposed by Heyd-Scuseria-Ernzerhof [37–39] with the exact exchange and generalized gradient exchange mixed in the ratio 1:3 is employed to treat the inter-atomic potentials. The SHF method is a more valid method for optical properties of materials as it overcomes the implicit self-interaction error of density-based exchange-correlation functionals and predicts reliable band gap values[40]. The ion-electron interaction was treated with the projector augmented wave (PAW) method [41]. Kinetic energy cut-off for the plane waves was set to 500 eV. The geometrical optimization was performed with no constraints on the unit cell vectors and atomic coordinates with the Perdew-Burke-Erzernhof parametrized generalized gradient approximation (GGA) functional which is sufficiently accurate for structural properties [42]. A high-density *k*-points grid was used to span the reciprocal space. The total energy tolerance was set to $10^{-7}$ eV.

## 3. Structure Determination

The Inorganic Crystal Structure Database (ICSD) [43] and American Mineralogist Crystal Structure Database (AMCSD) [44] are used for the functionality-based materials selection. These two databases together contain several hundred thousand of practically known compounds from inorganic, organo-inorganic and mineralogy fields. The library of compounds was filtered with additive search strings as zinc and carbon – carbon being the fundamental component of an organic-block. Other strings like hydrogen, oxygen, nitrogen, and phosphorus were also included in the search to match the scope of organic molecules common to the living matter. We identified thirteen distinct compounds that can reasonably be classified as Zn-based HOI materials. The compounds are named as samples Z3–Z15 and are summarized in Table 1, wherein we also list the functional group and the crystal structure. The wurtzite ZnO and zinc peroxide ($ZnO_2$) inorganic binary oxides of Zn, denoted as Z1 and Z2, respectively, are used as references to compare the properties of the cataloged compounds.

In preparing the catalog two difficulties were confronted that prohibited a straightforward use of crystal structures from the library dataset for first-principles computations. Firstly, the crystal structures of solid solution-type compounds are described by assigning multiple elements at common cation site with the composition accounted by proportional partial occupancy of the site. Practically, all the plane-wave DFT calculations require structural information with atomic sites assigned explicitly with individual elements from the periodic table. This necessitated transforming the composition of the models than those documented in the original database. The composition shift was maintained towards the Zn-rich side keeping in mind the motivation for Zn-based hybrid compounds. The second difficulty was realized with respect to the hydrogen atoms position, which are missing in the original database. It is well known that many spectroscopic techniques fall short in resolving the position of low-atomic number elements. This is either due to inadequacy of the characterization method or due to high-temperature measurements which dissolve the hydrogen-related spectra in the background noise. In our models the hydrogen atoms are supplemented to all the open bonds according to valence matching rules. The following models were geometrically optimized in order to ascertain that the local residual forces were minimized before any properties was calculated.

The organic building blocks for the Zn-based HOI oxides are found to be acceptor groups $[CO_3]^{2-}$, $[C_2O_4]^{2-}$, $[HCO_2]^-$, $[PO_4]^{3-}$, $[HCO_2]^-$, and $[N(CH_2HPO_3)_3]^{3-}$ or donor groups $[CH_3NH_3]^+$ and $[(CH_3)_2NH_2]^+$. It should be recalled that the $[CO_3]^{2-}$ is responsible for maintaining the pH balance of physiological reactions and $[PO_4]^{3-}$ ion is a functional moiety of hydroxyapatite which is a primary component of bones in vertebrates[45]. The phosphate group is also a major component of adenosine diphosphate and adenosine triphosphate that play an important role in the energy cycle in the living cells. These examples suggest a greater chance of sustaining Zn-HOI materials for bio-applications.

The crystal structures of Zn-based HOI materials in ball-stick representation are displayed in Fig. 1 (c)–(o). The binary zinc oxides, ZnO and $ZnO_2$ are shown Fig. 1 (a) and (b), respectively, for comparison. They show rich patterns with pure or mixed octahedral ($ZnO_6$) and tetrahedral ($ZnO_4$) motifs. The tetrahedron and octahedron motifs are shown as blue and gray polygons, respectively, in Fig. 1. The tetrahedron motif is common for inorganic zinc oxides, but octahedral motif is atypical and have been reported only in dispersed medium [46–51] and in other complex oxides [52,53]. It is known that compounds having highly electronegative components, such as $KZnF_3$ [54] and zinc peroxide ($ZnO_2$), stabilize octahedral motifs. The low packing density and bonding of the organic functional group to the vertex oxygen atom makes it possible to stabilize the octahedral $ZnO_6$ motif in HOI materials. Among the listed compounds, $((CH_3)_2NH_2)Zn_3(PO_4)(HPO_4)_2$ (Z5), $(CH_3NH_3)Zn_4(PO_4)_3$ (Z6), $Zn_8(HPO_4)_{16}(C_2H_8N)_8$ (Z10), $(Mg_2Zn)_8(CO_3)_2(OH)_6$ (Z12), and $Ca_3Zn_2(PO_4)CO_3(OH).2H_2O$ (Z14) have explicit $ZnO_4$ tetrahedral motifs and $CuZnO_2(CO_3)$ (Z3), $Zn(C_2O_4)$ (Z4), $Zn(N(CH_2PO_3H)_3)(H_2O)_3$ (Z7), $(CH_3NH_3)Zn(HCO_2)_3$ (Z8), and $Zn(CO_3)$ (Z15) have explicit $ZnO_6$ octahedral motifs. The compounds $Zn_4(CO_3)_2$ (Z9), $Zn_5(CO_3)_2$ (Z11) and $Zn_7(CO_3)_2(OH)_{10}$ (Z13) have mixed motifs with the number of tetrahedral to octahedral motifs in the ratio 1:4, 2:3, and 3:4, respectively.

The structural analysis carried out herein is thus useful to establish a classification scheme based on the local coordination of primary cation which has been applied for determination of certain physical properties[55]. For example, the arrangement of O-tetrahedra in zeolite ($SiO_4$) is central to improving the efficiency of their catalytic properties [56,57]. The three-dimensional network of octahedra and tetrahedra provide the pathways for ion diffusion in electrolytes for all-solid-state lithium-ion batteries [58,59]. Ferroelectricity, electrocaloric response and pyroelectric properties observed in Ti-based perovskites have typical $TiO_6$ octahedral motifs in their crystal structure [60] which display rich functional properties due to displacements and rotations of these octahedra. In superconducting materials, the correlation between the local structural motif, electronic configuration, and the critical temperature for the superconductor transition have been reported [61]. Furthermore, in the topic of point defects in solids, the local coordination of impurity atom plays a vital role in influencing the electronic structure and thus the conductivity of the system [62]. Computational determination of the structure-property relations of Zn-HOI materials for bio-applications is quite challenging but such an effort would have significant impact if supplemented by extensive experimental research.

## 4. Electronic Structure and Band Gap

The electronic structures of the compounds are calculated using the SHF method. To establish a baseline, we note that the experimental band gap of ZnO is reproduced by taking the exact exchange and GGA in the ratio 1:2 [63]. In the present calculations the ratio of 1:3 is retained as suggested in the rational for the SHF method [38] that allows to make comparison of results consistent across all the compounds. The optimum ratio can be decided by matching to experimental results, which for most of the compounds is sparsely available. The results are still applicable because varying the exchange ratio would vary the band gap values, but the band dispersion and the pattern of density of states (DOS) would remain similar. Results show that among the list of thirteen Zn-based HOI compounds, $CuZnO_2(CO_3)$ (Z3) and $Zn_8(HPO_4)_{16}(C_2H_8N)_8$ (Z10) are metallic while rest are semiconducting with wide band gap. The numerical value of band gap of each compound is listed in Table 1. Four compounds with distinct features are taken to compare the electronic DOS. The distinct features are: $ZnCO_3$ (Z15) and $Zn(C_2O_4)$ (Z4) vary by the type of functional groups, i.e. $[CO_3]^{2-}$ and $[C_4O_2]^{2-}$, respectively, and $Zn_5(CO_3)_2$ (Z11) and $Zn_7(CO_3)_2(OH)_{10}$ (Z13) vary by the number of tetrahedral and octahedral motifs in their crystal structures. The band gap of Z15, Z4, Z11 and Z13 are found to be 5.7 eV, 5.0 eV, 4.9 eV and 4.9 eV, respectively, from a detailed band structure analysis. The DOS diagrams for Z15, Z4, Z11 and Z13 are shown in Fig. 2(a)–(d), respectively. The total DOS is compared with the DOS of Zn atoms with octahedral coordination ($Zn_{oct}$) and tetrahedral coordination ($Zn_{tet}$), O atoms of from octahedral ($O_{oct}$) and tetrahedral ($O_{tet}$) motifs, and O atoms common to both tetrahedral and octahedral ($O_{tet-oct}$) motifs. The scale of DOS is adjusted in order to bring out the hybridization features more prominently. The zero in the energy axis refers to the Fermi level. The occupied Zn $d$-bands have majority of DOS weight in the energy range –8 eV and –4 eV. A small fraction of occupied Zn states spread across the valence band, which contain contributions mostly from O atoms. The peak of carbon at –9 eV is attributed primarily to the C–O bond from $CO_3$ and appears in the DOS of $ZnCO_3$ (Z15), $Zn_5(CO_3)_2$ (Z11) and $(Zn_7(CO_3)_2(OH)_{10})$ (Z13). The lower energy of C–O bonding state for $CO_3$ suggests that carbon-oxygen hybridization is relatively stronger in $CO_3$-based HOI compounds that containing $C_2O_4$ the group. The antibonding state of $CO_3$ is 7 eV above the Fermi level, while the antibonding state of $C_2O_4$ falls within the band gap region. The DOS describes the nature of chemical hybridization among the components of a material and helps decoding the charge redistribution associated with a structural change as in a reaction, which for this class of materials is yet to be investigated.

## 5. Optical Properties

The value of band gap and the orbital composition of the transition bands determine the optical properties. In order to compute optical properties, the frequency dependent dielectric functions of the compounds were determined by the linear response (density functional perturbation) theory [64,65]. The dielectric function, defined through $\epsilon = \epsilon_1 + i\epsilon_2$, is a direction dependent quantity consisting of a real part ($\epsilon_1$) which represents dispersion of light in the medium and an imaginary part ($\epsilon_2$) which describes the dissipation of energy into the medium. The absorption coefficient $\alpha(\omega)$ and the refractive index $n(\omega)$ as functions of the frequency $\omega$ are calculated from the following relations:

$$\alpha(\omega) = \omega\sqrt{2\sqrt{\epsilon_1^2(\omega) + \epsilon_2^2(\omega)} - 2\epsilon_1(\omega)}, \text{ and} \qquad (1)$$

$$n(\omega) = \sqrt{\frac{\sqrt{\epsilon_1^2(\omega) + \epsilon_2^2(\omega)} + \epsilon_1(\omega)}{2}}. \qquad (2)$$

We focus here on the isotropic magnitude of $\epsilon$ by averaging the Cartesian components of the real ($\epsilon_1$) and the imaginary ($\epsilon_2$) parts separately, which are then used to calculate the adsorption and refractive index. As an example, we provide here results for compound Z6, $(CH_3NH_3)Zn_4(PO_4)$. The crystal structure of Z6 and the corresponding DOS are shown in Figs. 3(a) and 3(b), respectively. The optical excitation in semiconductors are from the occupied valency band to the unoccupied conduction band. Light frequencies below the band gap energy is not absorbed. The absorption coefficient and the refractive index of Z6 as a function of energy is shown in Fig. 3(c).

The optical absorption coefficient is large in the energy range of 7–20 eV implying a blocking capacity for UV radiation of the same energy range.

A similar analysis for other Zn-HOI compounds were carried out. The optical absorption coefficients and refractive indices for all the compounds with non-zero band gap as a function of energy are plotted in Fig. 4(a) and (b), respectively. The energy range is restricted to the UV spectrum of only 3.2–12.4 eV with a view of potential applications related to UV lasers. The UV spectrum is generally sub-classified into three energy regions, the near UV (NUV) with energy (wavelength) in the range 3.10–4.13 eV (400–300 nm), the mid-UV (MUV) for 4.13–6.20 eV (300–200 nm), and the far UV (FUV) for 6.20–12.40 eV (200–122 nm). These ranges are marked as brown, orange and yellow regions, respectively, in Fig. 4. The corresponding wavelength in nanometers is provided at the top horizontal axis. The optical adsorption spectra of Fig. 4(a) shows that all Zn-HOI compounds described herein have a relatively high absorbance capacity for FUV. This property could lead to applications such as UV coating of sensitive devices. But, the presence of Zn provides a compelling reason for potential use in the health sector and for bio-applications. In this regard, it is worth to review the effect of laser action on tissue which involves either one or a combination of important phenomena; photothermal (involves photovaporization and photocoagulation) effect, photochemical (photoradiation and photoablation) effect, and photoionizing effect. Just to give an idea with respect to potential applications in biomedical engineering and health care, we plot in Fig. 4(b) the refractive index of the compounds within the shaded horizontal region, 1.36 – 1.49, which corresponds to the range of refractive indices for mammalian tissues[66,67]. The refractive index of vacuum is marked as a dashed horizontal line.

Refractive index of Zn-HOI materials being larger than that of living tissues has a major implication. It adds an effective "lensing effect" and confines the light beam to a narrow region – similar to converging the sunlight to a point by using a convex lens. This is possible because light beam travelling from a rarer medium (lower refractive index, $n_1$) to a denser medium (higher refractive index, $n_2$) bends the beam towards the normal at the interface. The angle of bending is proportional to the ratio of the refractive index ($n_2/n_1$). This physical effect matters much in radiation therapy where securing the beam to a narrow confinement is of primary importance so that the scattered radiation does not excite the surrounding tissue. The absorption coefficient of Zn-HOI materials is three orders of magnitude larger than average tissue; a comparison is given in

Fig. 5. This implies that in an integrable environment most of the radiation energy would be absorbed by the Zn-HOI material. Depending on the requirements, this can be beneficial for protecting the underlying tissue from radiation damage or subjecting a targeted heat delivery to destroy the tissue as in selective photothermolysis therapy. The high UV absorption of these materials also make it attractive for UV detection and protection (such as skin care products, sunscreen lotions) applications.

## 6. Conclusions

Discovery and development of functional materials for targeted applications present a major challenge. In the search for Zn based HOI materials with high refractive indices, we considered compounds from two crystallographic repositories, ICSD and AMCSD. We successfully identified thirteen Zn-based HOI compounds and investigated their electronic structure and optical properties with DFT. The results show that these compounds display relatively wide band gaps and have a propensity for large absorption of light in the MUV and FUV range. The high refractive index values of the materials identified here would enable the use of these compounds as a protective coating against UV radiation. The presence of non-toxic Zn and organic groups $CH_3NH_3$, $CO_3^{2-}$, $C_2O_4^{2-}$ and $PO_4^{3-}$ which play prominent roles in metabolic activities point towards applications in areas such as life sciences and medicine. While only thirteen Zn-based HOI compounds are identified in the present study, the list can be expanded with targeted experimental work supplemented with computational tools developed herein.

**Author Contributions**

TS conceived the original idea. SKN and TS performed the DFT calculations and lead the analysis efforts. TS, UAA, SPP and SKN contributed in preparing the manuscript.

**Additional Information**

**Competing Interests:** The authors declare no competing interests.

**Data Availability:** The raw/processed data required to reproduce these findings cannot be shared at this time as the data also forms part of an ongoing study.

**Figure Captions**

**Figure 1**  Crystal structure of Zn-based compounds Z1-Z15 is shown in (a)-(o). The chemical formula and band gap of each of the compounds is provided in Table 1. The $ZnO_4$ tetrahedral motif and $ZnO_6$ octahedral motifs are shown as blue and gray colored polygons, respectively.

**Figure 2**  Density of states for (a) $Zn_5(CO_3)_2$ (Z11), (b) $Zn(C_2O_4)$ (Z4), (c) $Zn_7(CO_3)_2(OH)_{10}$ (Z13), and (d) $ZnCO_3$ (Z15) calculated using the SHF method.

**Figure 3**  (a) Crystal structure, (b) electronic DOS, and (c) absorption coefficient and refractive index of $(CH_3NH_3)Zn_4(PO_4)$ (Z6). The Fermi level is set to zero for the DOS.

**Figure 4**  The (a) optical adsorption coefficient and (b) refractive index for HOI samples with finite band gap (see Table 1) as a function of energy in the UV spectrum. The refractive index of mammalian tissues falls in the range 1.36–1.49 shown as solid red lines.

**Figure 5**  The absorption coefficient of Zn-HOI materials compared to the absorption spectra of proteins, hemoglobin, melanin, water and tissue. Zn-HOI materials is three times larger in magnitude than average tissues absorption of UV light, ensuring that selectivity of laser to tissue could occur. Part of the data adopted from Ref.[68].

**Table I.** Chemical compositions of the Zn-based HOI compounds. Atomic coordinate source, organic functional group(s), space group in Hermann-Mauguin (H-M) notation, corresponding mineral and band gap (eV) values obtained from SHF method are also shown.

| Sample | Compound | Source | Functional group | Space group | Band gap |
|---|---|---|---|---|---|
| Z1 | ZnO | - | - | $P6_3mc$ | 2.4 |
| Z2 | $ZnO_2$ | - | - | $Pa\text{--}3$ | 4.3 |
| Z3 | $CuZnO_2(CO_3)$ | ICSD-109166, AMCSD-0011132 | $[CO_3]^{2-}$ | $P2_1/c$ | 0.0 |
| Z4 | $Zn(C_2O_4)$ | ICSD-109665, AMCSD-0019967 | $[C_2O_4]^{2-}$ | $C2/c$ | 5.0 |
| Z5 | $((CH_3)_2NH_2)Zn_3(PO_4)(HPO_4)_2$ | ICSD-110594 | $[(CH_3)_2NH_2]^+$, $[HPO_4]^{2-}$ | $P2_1/c$ | 5.9 |
| Z6 | $(CH_3NH_3)Zn_4(PO_4)_3$ | ICSD-170961 | $[CH_3NH_3]^+$, $[PO_4]^{3-}$ | $Pbca$ | 5.6 |
| Z7 | $Zn(N(CH_2PO_3H)_3)(H_2O)_3$ | ICSD-172319 | $[N(CH_2HPO_3)_3]^{3-}$ | $P2_1/c$ | 2.8 |
| Z8 | $(CH_3NH_3)Zn(HCO_2)_3$ | ICSD-186753 | $[CH_3NH_3]^+$, $[HCO_2]^-$ | $Pnma$ | 6.2 |
| Z9 | $Zn_4(CO_3)_2$ | AMCSD-0009883 | $[CO_3]^{2-}$ | $P2_1/m$ | 0.0 |
| Z10 | $Zn_8(HPO_4)_{16}(C_2H_8N)_8$ | AMCSD -0012623 | $[(CH_3)_2NH_2]^+$, $[PO_4]^{3-}$ | $Cc$ | 6.9 |
| Z11 | $Zn_5(CO_3)_2$ | AMCSD-0009288 | $[CO_3]^{2-}$ | $C2/m$ | 4.9 |
| Z12 | $(Mg_2Zn)_8(CO_3)_2(OH)_6$ | AMCSD-0009744 | $[CO_3]^{2-}$ | $P1$ | 3.4 |
| Z13 | $Zn_7(CO_3)_2(OH)_{10}$ | AMCSD-0001284 | $[CO_3]^{2-}$ | $P\text{--}1$ | 4.9 |
| Z14 | $Ca_3Zn_2(PO_4)CO_3(OH).2H_2O$ | AMCSD-0007259 | $[CO_3]^{2-}$, $[PO_4]^{3-}$ | $C2/c$ | 6.3 |
| Z15 | $Zn(CO_3)$ | AMCSD-0000102 | $[CO_3]^{2-}$ | $R\text{--}3c$ | 5.7 |

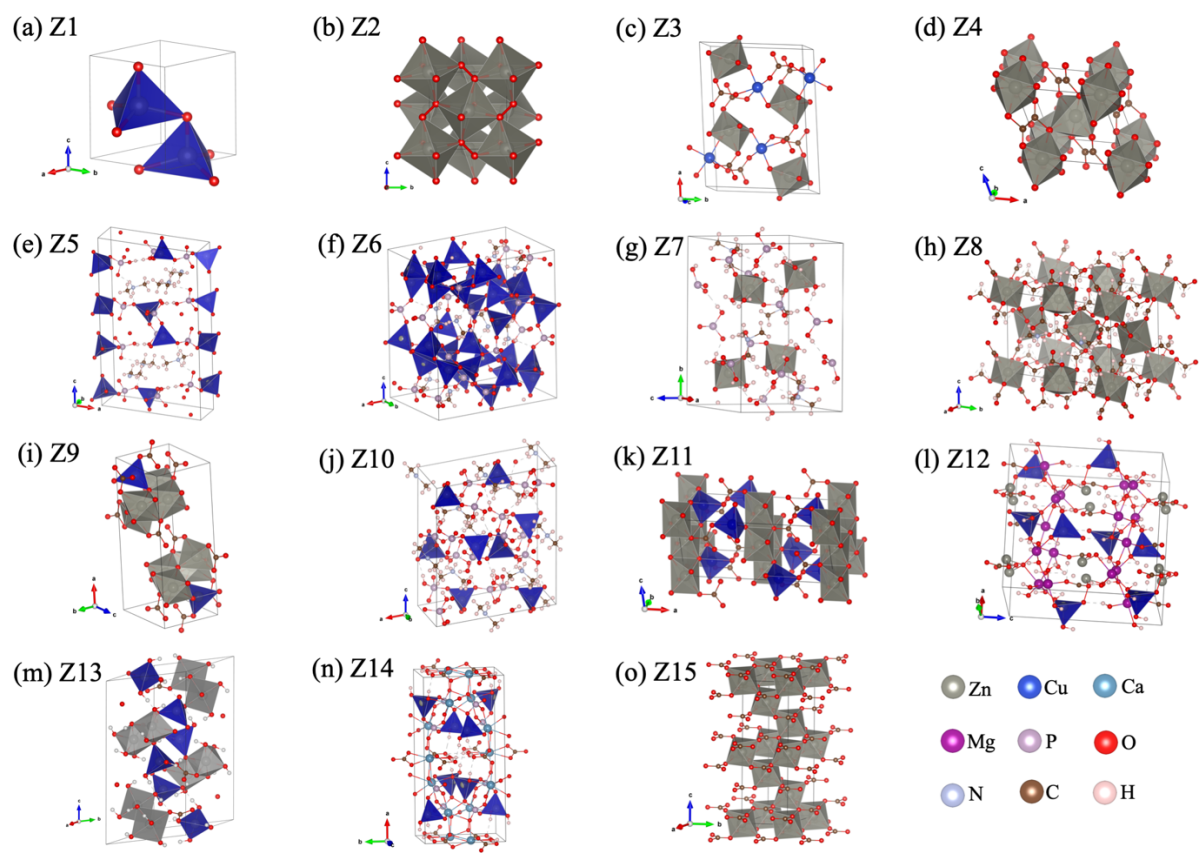

**Figure 1**

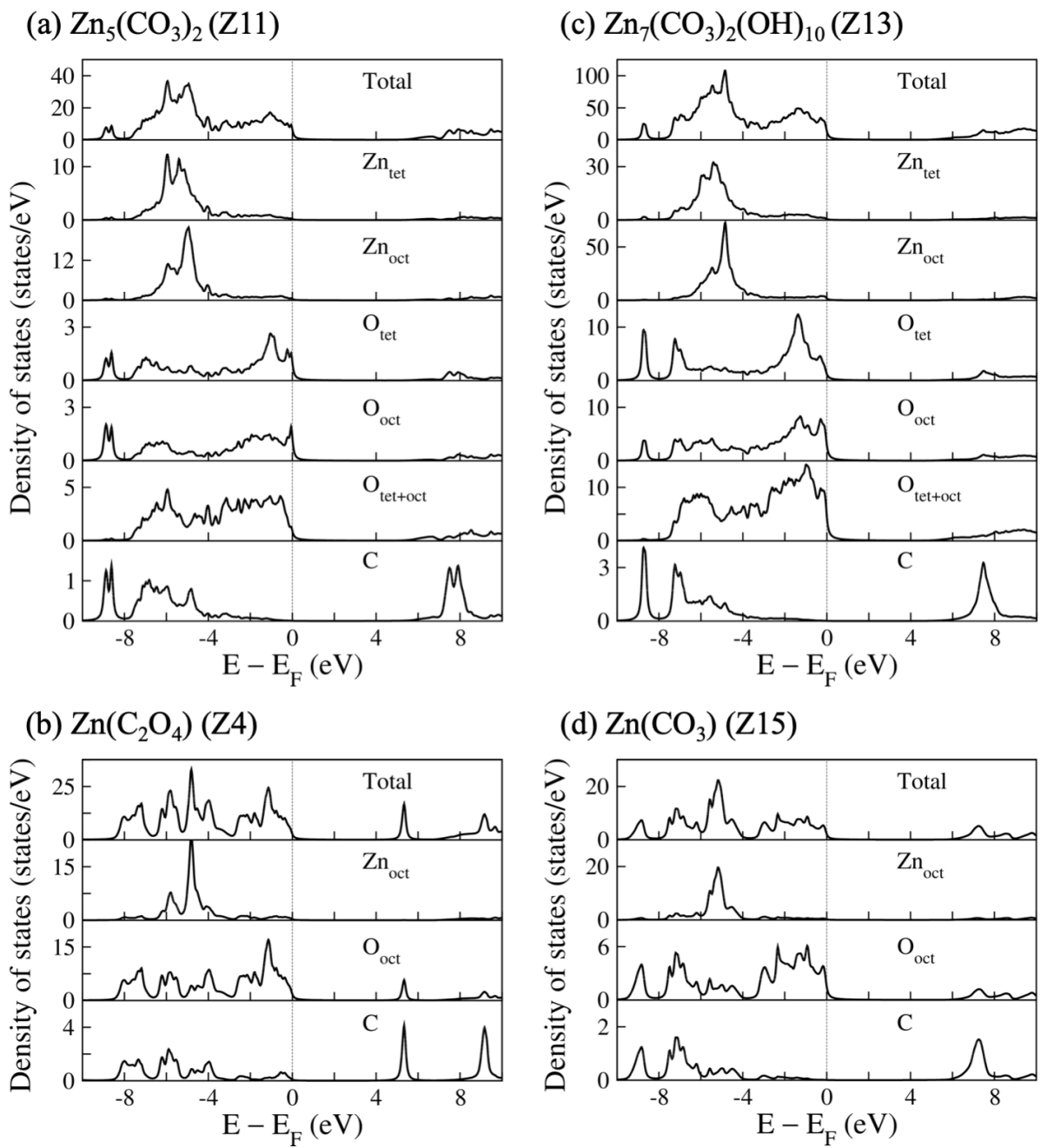

**Figure 2**

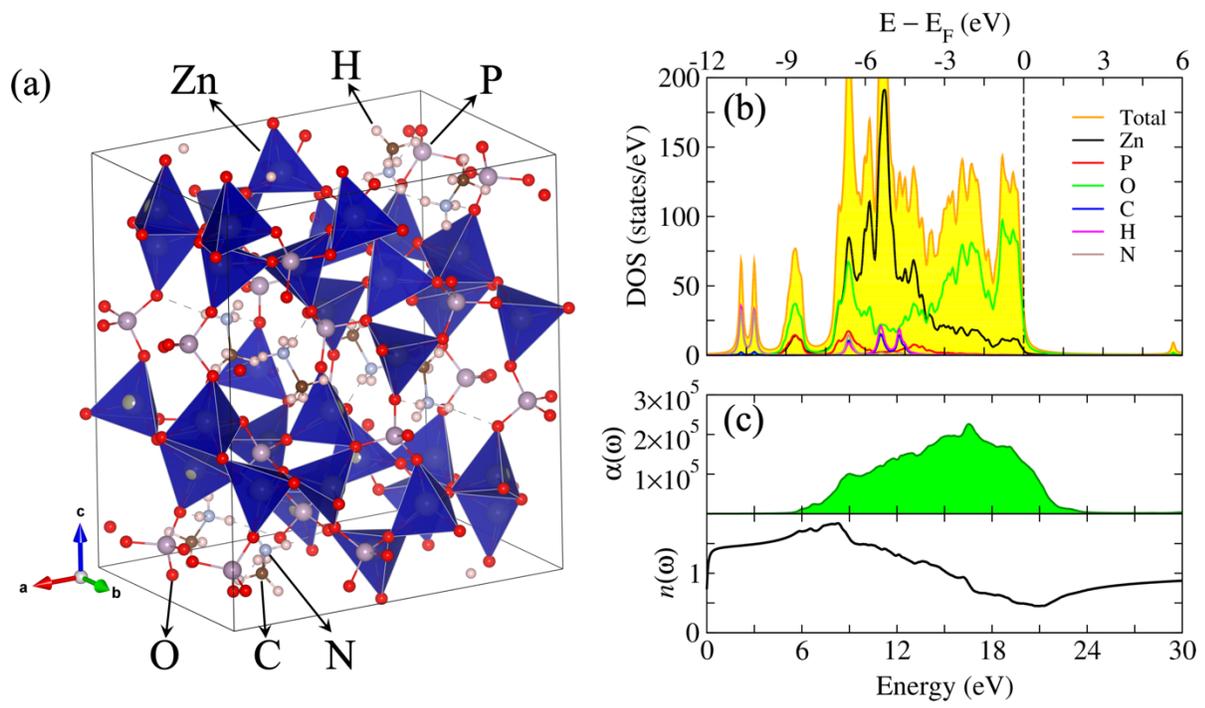

**Figure 3**

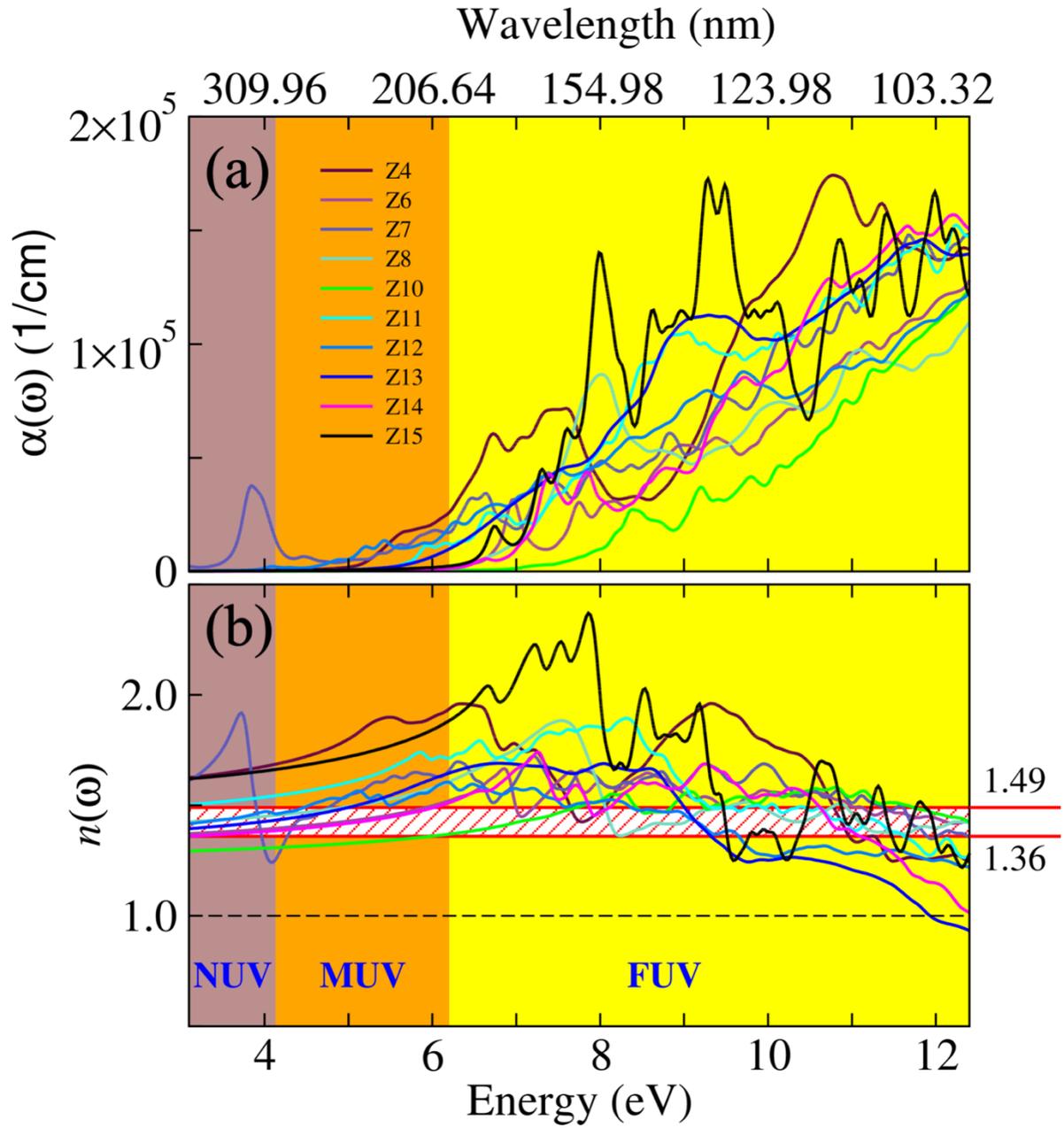

**Figure 4**

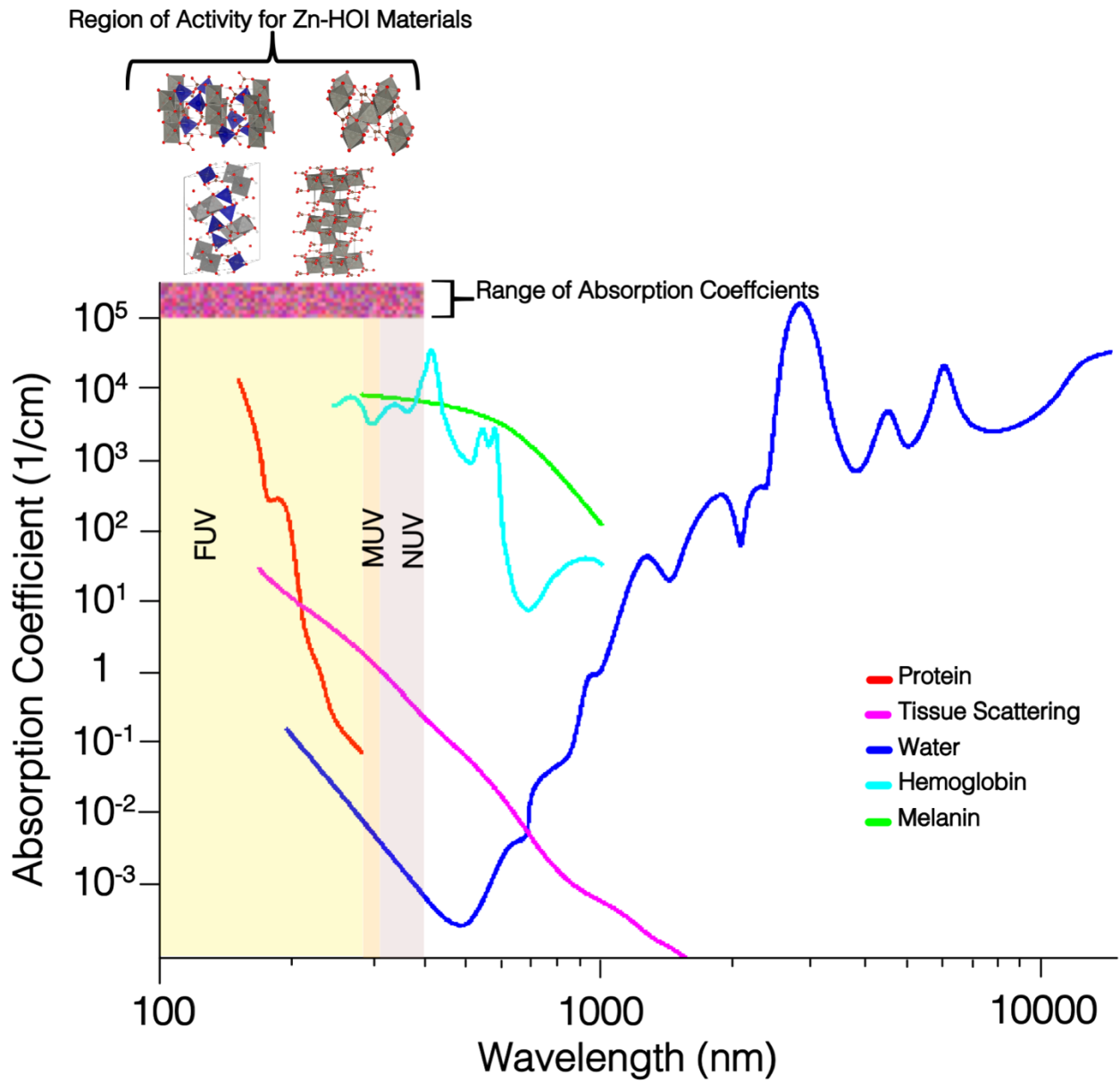

**Figure 5**